\documentclass[12pt]{article}
\usepackage{epsf}
\usepackage[dvips]{graphicx,color}

\renewcommand{\thefootnote}{\#\arabic{footnote}}
\begin{document}

\newcommand{\gtrsim}{ \mathop{}_{\textstyle \sim}^{\textstyle >} }
\newcommand{\lesssim}{ \mathop{}_{\textstyle \sim}^{\textstyle <} }

\renewcommand{\thefootnote}{\fnsymbol{footnote}}
\setcounter{footnote}{0}
\begin{titlepage}

\def\thefootnote{\fnsymbol{footnote}}

\begin{center}

\hfill TU-643\\
\hfill hep-ph/0112335\\
\hfill December, 2001\\

\vskip .5in

{\Large \bf
  Cosmic Microwave Background Anisotropy \\
with Late Time Entropy Production\footnote{
  Talk given at the 5th RESCEU International Symposium on
``New Trends in Theoretical and Observational Cosmology,''
Tokyo, November 2001. }
}

\vskip .45in

{\large
  Takeo Moroi and Tomo Takahashi
}

\vskip .45in

{\em
  Department of Physics,  Tohoku University, Sendai 980-8578, Japan
}

\end{center}

\vskip .4in

\begin{abstract}
We discuss effects of cosmological moduli fields on
the cosmic microwave background (CMB).  If a modulus field $\phi$ once
dominates the universe, the CMB we observe today is from the decay of
$\phi$ and its anisotropy is affected by the primordial fluctuation in
the amplitude of the modulus field.  As a result, constraints on the
inflaton potential from the CMB anisotropy can be relaxed.  In
addition, with the cosmological moduli fields, {\sl correlated}
mixture of adiabatic and isocurvature fluctuations may be generated,
which results in enhanced CMB angular power spectrum at higher
multipoles relative to that of lower ones.  Such an enhancement can be
an evidence of late time entropy production due to the cosmological
moduli fields, and may be observed at on-going and future experiments.
\end{abstract}
\end{titlepage}

\renewcommand{\thepage}{\arabic{page}}
\setcounter{page}{1}
\renewcommand{\thefootnote}{\#\arabic{footnote}}
\setcounter{footnote}{0}

\section{Introduction}
In superstring theory \cite{Polchinski}, it is well known that there
are various flat directions parameterized by scalar fields.
(Hereafter, we call these fields as ``moduli'' fields.)  Since their
potential is usually generated by effects of supersymmetry (SUSY)
breaking, their masses are expected to be of the order of the
gravitino mass.  Although masses of the moduli fields can be as light
as (or even lighter than) the electroweak scale, moduli fields do not
affect collider experiments since their interactions are suppressed by
inverse powers of the gravitational scale.

Cosmologically, however, they may cause serious problems
\cite{PRL131-59}.  If the mass of the moduli fields
$m_\phi\mathop{}_{\textstyle \sim}^{\textstyle <} O(10\ {\rm TeV})$,
reheating temperature of the universe becomes lower than $\sim 1\ {\rm
  MeV}$. With such a low reheating temperature, the success of the
standard big-bang nucleosynthesis (BBN) is spoiled. For lighter moduli
fields ($m_\phi \mathop{}_{\textstyle \sim}^{\textstyle <} O(100\ {\rm
  MeV})$), they survive until today and overclose the universe.  One
solution to these difficulties is to push up the mass of the moduli
fields \cite{Heavyphi}.  In particular, in Ref.\ \cite{NPB570-455}, it
was pointed out that the scenario with heavy moduli can naturally fit
into the framework of the anomaly-mediated SUSY breaking \cite{AMSB}.
Indeed the reheating temperature can be higher than $\sim 1\ {\rm
  MeV}$ if $m_\phi \mathop{}_{\textstyle \sim}^{\textstyle >}O(10\ 
{\rm TeV})$.  In this case, the BBN occurs after the decay of the
modulus field.  Although the thermal history after the BBN is mostly
the same as the standard one, cosmology before the modulus decay is
completely different.  Importantly, fluctuations of moduli fields
affect the cosmic microwave background (CMB).  In this talk, we
consider this scenario and study its consequence in the CMB.

\section{Effects of moduli fields on CMB}
Now we discuss the CMB angular power spectrum from the scenario with
the cosmological modulus field \cite{Moroi_Takahashi_2001}. First, we
emphasize that there are two independent sources of the CMB
anisotropy, i.e., primordial metric perturbation $\Psi_{\rm i}$
induced by inflaton field fluctuation and amplitude fluctuation of the
modulus field $\delta\phi_{\rm i}$ which is of order $H_{\rm
  inf}/2\pi$ where $H_{\rm inf}$ is the Hubble parameter during
inflation.  Therefore, the resultant CMB angular power spectrum $C_l$
is given in the following form:
\begin{equation}
    C_l = C_l^{\rm (adi)} + C_l^{(\delta\phi)}.
\end{equation}
Here, $C_l^{\rm (adi)}$ is from the perturbation in the inflaton
field, which is of order $\Psi_{\rm i}^2$.  On the contrary,
$C_l^{(\delta\phi)}$ is from the primordial fluctuation of the modulus
amplitude, which is of order $\delta\phi_{\rm i}^2$.  (Notice that
there is no term which is of order $\Psi_{\rm i}\delta\phi_{\rm i}$
since two fluctuations are uncorrelated.)  $C_l^{\rm (adi)}$ can be
calculated by following the standard method.  In calculating
$C_l^{(\delta\phi)}$, we must specify the origin of CDM and baryon.
Here, we assume that CDM is generated by the decay of $\phi$.  In this
case, after the decay of the modulus field, there is no entropy
between CDM and radiation.  On the contrary, we consider two
possibilities of generating baryon asymmetry: (i) the baryon asymmetry
is (somehow) generated at the time of (or after) the decay of $\phi$,
or (ii) the Affleck-Dine (AD) mechanism \cite{NPB249-361} generates
the baryon number.

Let us first consider the case (i).  In this case, there is no entropy
between baryon and radiation, so the cosmic fluctuations are
same as the conventional adiabatic case once the modulus field decays.
Thus, if we neglect the scale dependences of $\Psi_{\rm i}$ and
$\delta\phi_{\rm i}$, $C_l^{(\delta\phi)}$ is proportional to
$C_l^{\rm (adi)}$.  In this case, the CMB angular power spectrum is
the same as the usual adiabatic case if the normalization of the
initial fluctuations are properly chosen.  However, this fact has
significant implications when we construct a model of inflation. First
of all, since curvature perturbation induced by the modulus field
$\Psi^{(\delta\phi)}$ is of order $H_{\rm inf}/\bar{\phi}_{\rm i}$
where $\bar{\phi_{\rm i}}$ is the initial amplitude of $\phi$, 
large cosmic perturbation can be generated by lowering
$\bar{\phi}_{\rm i}$ even if $H_{\rm inf}$ is small.  Furthermore,
usually, scale dependence of $\delta\phi_{\rm i}$ is milder than that
of $\Psi_{\rm i}$.  Thus, when $C_l^{\rm (adi)}\ll C_l^{(\delta\phi)}$
is realized, the resultant CMB angular power spectrum may be like that
from the scale-invariant adiabatic perturbation even if $\Psi_{\rm i}$
has a strong scale dependence.  These facts relax constraints on the
potential of the inflaton field \cite{Moroi_Takahashi_2001,inf_const}.

Next we consider the case (ii).  Here we assume that the initial value
of the fluctuation in the AD field is negligibly small.\footnote{
This may happen when, for example, the
effective mass of the AD field is comparable to $H_{\rm inf}$
during the inflation. }
\begin{figure}[t]
  \begin{center}
    \includegraphics[height=13pc]{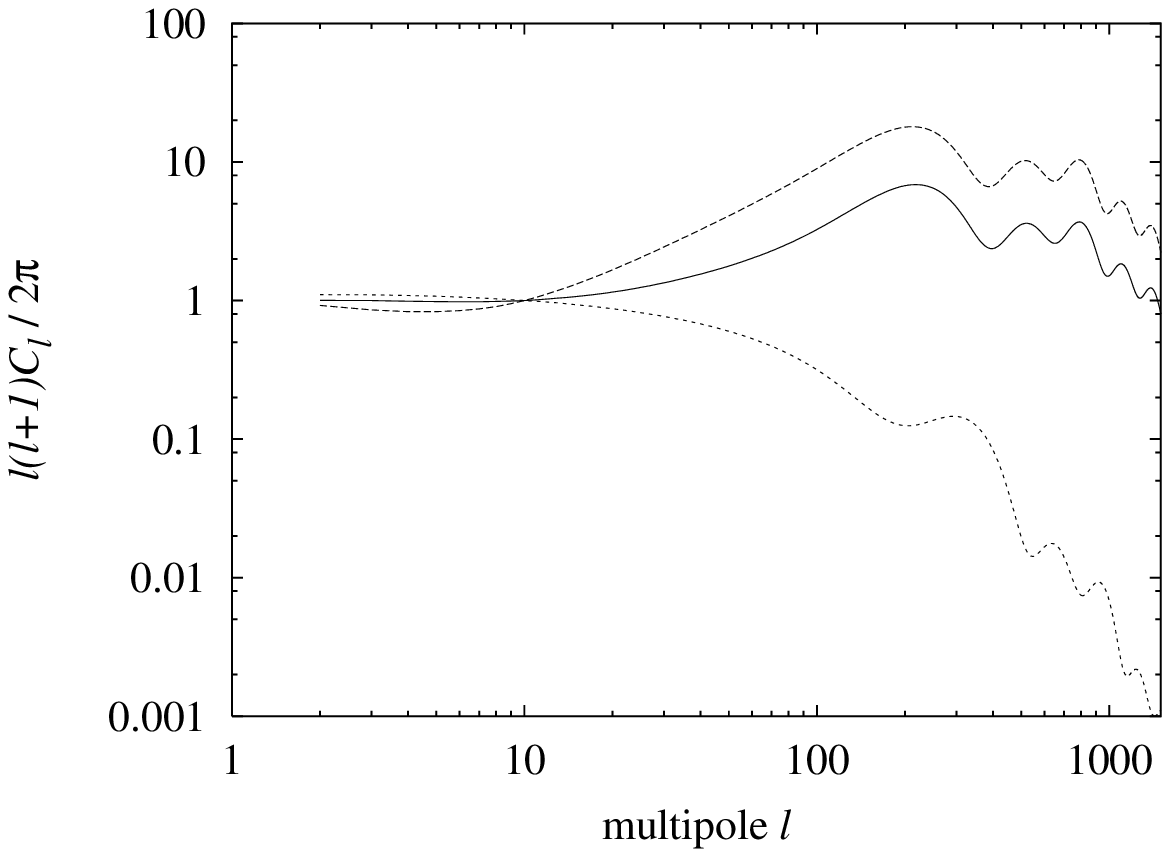}\includegraphics[height=13pc]{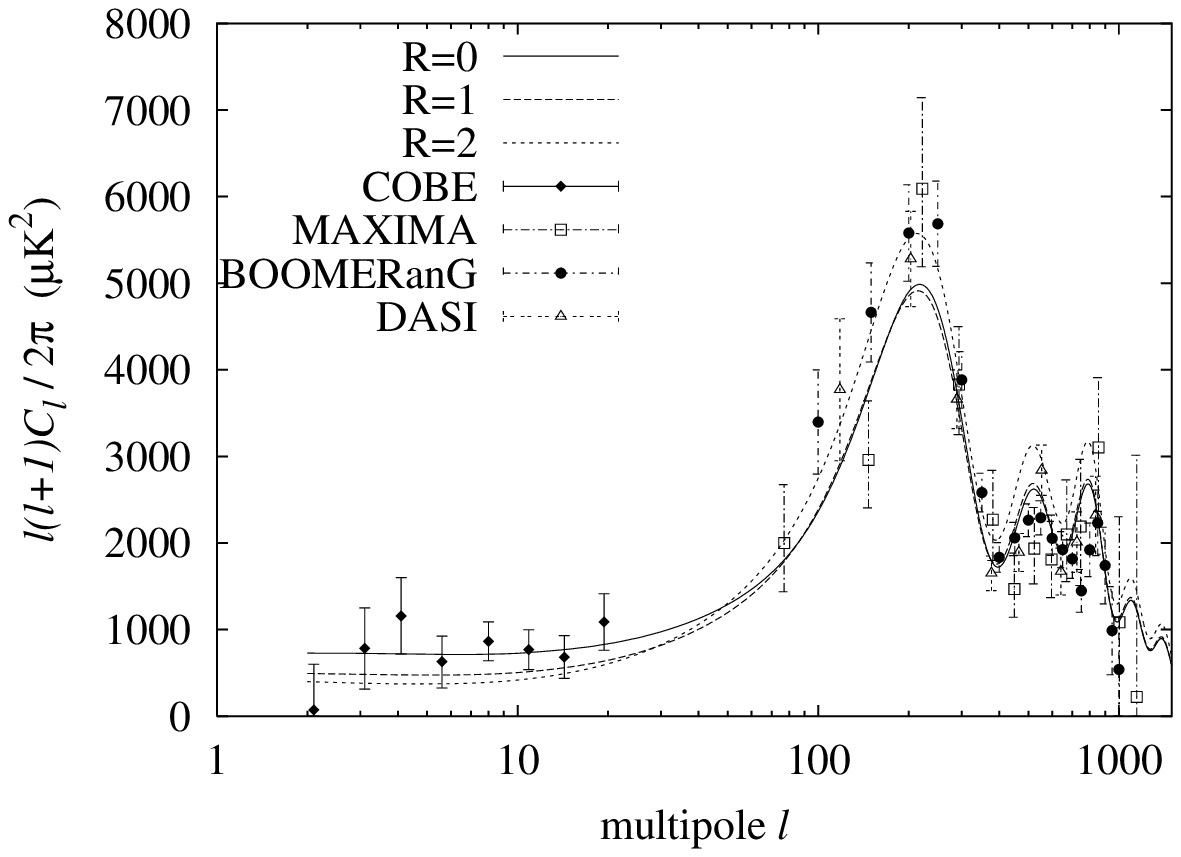}
  \end{center}
  \caption{{\it Left}: The CMB angular power spectrum $C_l^{(\delta\phi)}$ 
    for the case with the correlated isocurvature perturbation in the
    baryonic sector (dashed line), as well as $C_l$ for purely
    adiabatic (solid line) and purely baryonic isocurvature (dotted
    line) cases.  We used the normalization
    $[l(l+1)C_{l}/2\pi]_{l=10}=1$.  {\it Right}: The CMB angular power
    spectrum $C_l$ for $R=0$ (solid), $R=1$ (dashed), and $R=2$
    (dotted) where $R \equiv S_{\rm i}/\Psi^{\rm (adi)}$.  The overall
    normalization of $C_l$ is determined to be best fitted to the the
    observational data \cite{data}. In these figures, the cosmological
    parameters are taken such that $h=0.65$, $\Omega_{\rm
      b}h^2=0.019$, $\Omega_{\rm m}=0.4, \Omega_{\Lambda}=0.6$, and
    scale-invariance is assumed both for $\Psi_{\rm i}$ and $S_{\rm
      i}$.}
  \label{fig:cmb}
\end{figure}
Even when there is no primordial fluctuation in the AD field, however,
non-vanishing isocurvature fluctuation is generated in the
baryonic sector if $\delta\phi_{\rm i}\neq 0$.  Since, initially,
there is no fluctuation in the baryon energy density, the entropy
between baryon and $\phi$ exists; $S_{b\phi} \equiv \delta_b -
\delta_\phi = S_{\rm i} \ne 0,$ where $\delta_x \equiv
\delta\rho_x/\rho_x.$ This entropy is conserved until the
modulus field decays, and it becomes the entropy between photon
and baryon after the decay of $\phi$.  Therefore, in
calculating $C_l^{(\delta\phi)}$ which is generated by the primordial
fluctuation in the modulus amplitude, initial condition for the
baryonic density fluctuation is different from the conventional
adiabatic one, and is given by, in deep radiation dominated epoch 
\cite{Moroi_Takahashi_2001},
\begin{eqnarray}
    \delta_b  
    = S_{\rm i} + (3/4)
    \delta_\gamma
    = 4.5 \Psi^{(\delta\phi)} 
    + (3/4)
    \delta_\gamma.
    \label{db_i}
\end{eqnarray}
Importantly, the curvature perturbation $\Psi^{(\delta\phi)}$ is
induced by the initial entropy perturbation $S_{\rm i}$. In addition,
notice that the first term in the right-handed side of the above
equation does not exist in the usual adiabatic ones.  Initial
conditions for other perturbations are the same as the usual adiabatic
case.  The first term in Eq.\ (\ref{db_i}) gives rise to the
non-vanishing entropy in the baryonic sector and hence we call it as
the ``isocurvature'' term.  Here, it should be emphasized that such an
isocurvature fluctuation is {\sl correlated} with the contribution
from $\Psi^{(\delta\phi)}$ which gives rise to the effect like the
conventional adiabatic fluctuation.  Thus, the effect of this
``isocurvature'' fluctuation is completely different from the
conventional {\sl uncorrelated} isocurvature fluctuation.

In Fig.\ \ref{fig:cmb}, we show the total CMB angular power spectrum
with such a correlated isocurvature fluctuation in the baryonic
density fluctuation.  Since there are two sources of the cosmic
perturbations, $\Psi_{\rm i}$ and $S_{\rm i}$ (or equivalently,
$\delta\phi_{\rm i}$), we define $R \equiv S_{\rm i}/\Psi^{\rm (adi)},$
where $\Psi^{\rm (adi)}$ is the gravitational potential in the
adiabatic mode at the radiation dominated epoch after the modulus
decay.  (For simplicity, we neglect the scale dependence of $R$.)
Treating $R$ as a free parameter, we plot the total $C_l$ for several
values of $R$ in Fig.\ \ref{fig:cmb}.  As one can see, correlated
isocurvature perturbation in the baryonic sector may result in an
enhancement of $C_l$ at higher multipoles relative to that at lower
ones.  Therefore, on-going and future experiments will give us
interesting tests of the scenario with the cosmological moduli fields.

\section{Conclusion}
We have studied the effects of the cosmological moduli fields on the
CMB anisotropy.  In the scenario with the cosmological moduli fields,
{\sl correlated} isocurvature fluctuation may exist in the baryonic
sector which results in enhanced CMB angular power spectrum at higher
multipoles relative to that at lower ones.  In addition, even in the
case where there is no isocurvature perturbation, the cosmological
modulus field may have important implication to the model-building of
the inflation.



\end{document}